\documentclass[lettersize,journal]{IEEEtran}
\usepackage{amsmath,amsfonts}
\usepackage{algorithmic}
\usepackage{algorithm}
\usepackage{array}
\usepackage[caption=false,font=normalsize,labelfont=sf,textfont=sf]{subfig}
\usepackage{textcomp}
\usepackage{stfloats}
\usepackage{url}
\usepackage{verbatim}
\usepackage{graphicx}
\usepackage{cite}
\usepackage{bm}

\usepackage{amsmath}
\usepackage{amssymb}

\begin{document}

\title{Enabling High-Bandwidth Flash for Generative Recommendation Serving with Write-Aware KV Cache Policy}

\author{Danni Peng, Kai Wu, Tianyu Zuo, Pengfei Xia, and Hui Zang

\thanks{This work has been submitted to the IEEE for possible publication. Copyright may be transferred without notice, after which this version may no longer be accessible.\

The authors are with Huawei Technologies Co., Ltd. Corresponding author's e-mail: peng.danni@huawei.com.}
}

\markboth{Preprint}
{Shell \MakeLowercase{\textit{et al.}}: A Sample Article Using IEEEtran.cls for IEEE Journals}


\maketitle
\begin{abstract}
Generative recommendation (GR) systems increasingly leverage user-level KV cache reuse to avoid recomputing long user histories. However, the growing KV cache capacity and bandwidth requirements introduce new challenges for memory system. High-Bandwidth Flash (HBF) provides a promising solution by offering substantially higher capacity than HBM while approaching HBM-class read bandwidth, enabling larger-scale KV cache retention and improved serving throughput. Yet conventional Least-Recently-Used (LRU) KV cache management tightly couples KV cache writes with cache misses, generating excessive write traffic that rapidly exhausts flash endurance. In this work, we evaluate a write-aware KV cache policy based on admission-controlled LRU-$K$ for HBF-based GR serving. By filtering low-reuse users before cache admission, LRU-$K$ decouples KV cache writes from misses and significantly reduces unnecessary writes. We develop an analytical model to characterize GR serving performance, KV cache write traffic, and HBF lifetime, and evaluate performance across diverse memory systems and GR workloads. Our results show that HBF-based systems achieve 3.8--4.7$\times$ higher throughput than HBM-only systems. Moreover, LRU-$K$ extends HBF lifetime from about one year under conventional LRU to over six years with a moderate $K=10$, while maintaining comparable or even slightly improved throughput. These results highlight the importance of write-aware KV cache policy for sustainable HBF-based GR serving.
\end{abstract}


\section{Introduction}
 \IEEEPARstart{R}{ecently}, generative recommendation (GR) has gained increasing attention due to its scaling potential.
 Built upon transformer-based architectures, GR models improve recommendation quality by scaling model capacity and incorporating longer user interaction histories \cite{zhai2024actions,zhou2025openonerec}. \

\begin{figure}[!t]
\centering
\includegraphics[width=0.9\columnwidth]{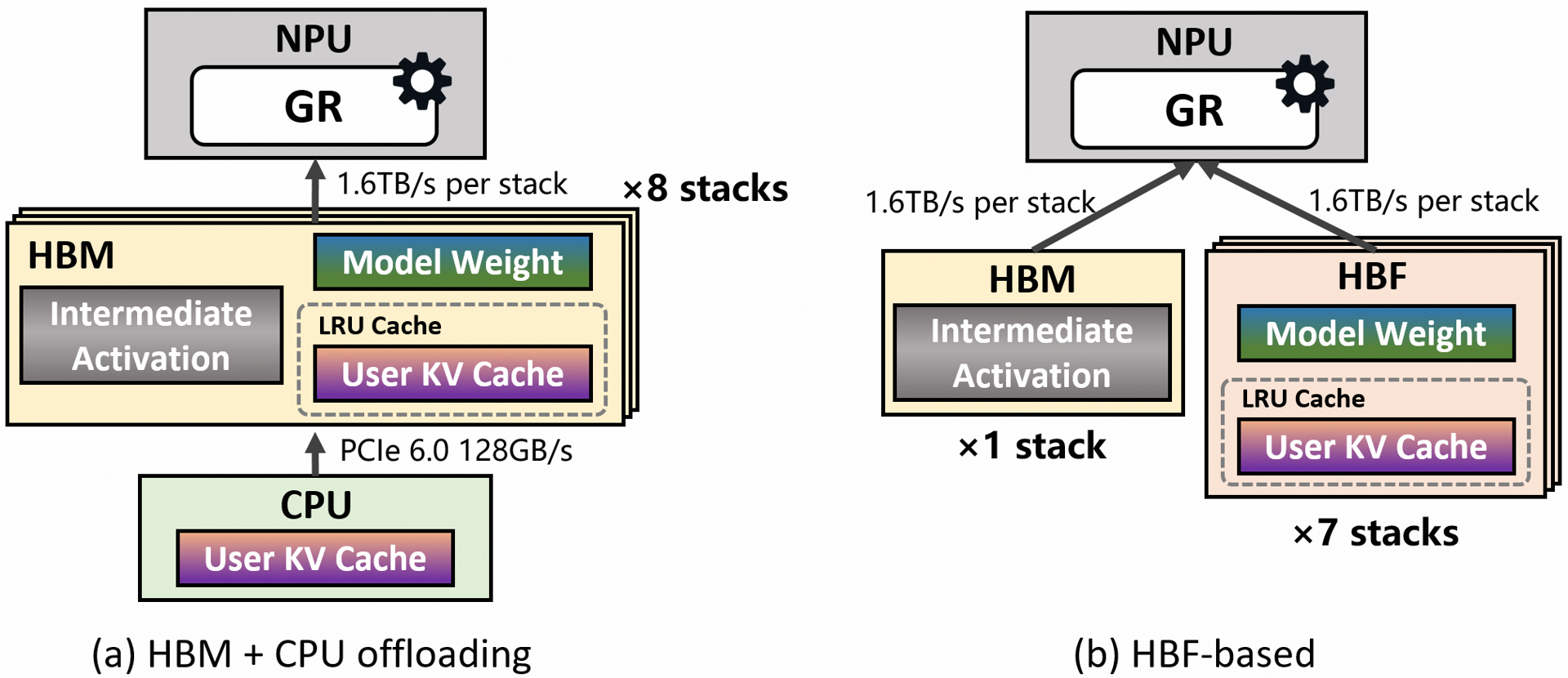}
\caption{GR serving on conventional HBM+CPU versus an HBF-based system}
\label{fig:archi}
\end{figure}

\begin{figure}[!t]
\centering
\includegraphics[width=\columnwidth]{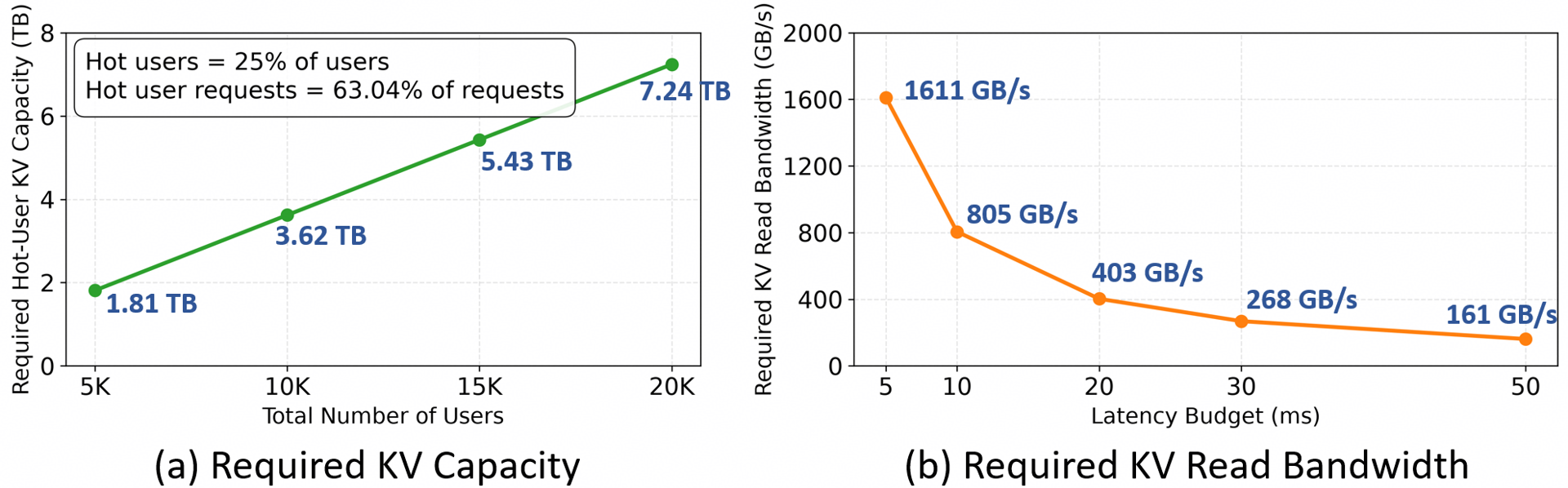}
\caption{Required KV Cache Capacity and Read Bandwidth for HSTU-10B.}
\label{fig:bw_cap}
\end{figure}

To reduce the latency overhead of recomputing long user histories, recent GR systems employ user-level KV cache reuse across serving requests. As shown in Figure~\ref{fig:archi}a, current industry practice commonly places user KV caches in accelerator memory, such as HBM, alongside model weights and intermediate activations to meet stringent latency requirements. CPU memory can further be leveraged to expand cache capacity by offloading inactive user KV \cite{wang2026mtserve,sun2026bat}. 
However, as GR services evolve toward larger models, broader user populations, and tighter latency SLOs, this design faces two key challenges. First, persistent user-level KV caches impose substantial storage demands that scale with the user base. As shown in Figure~\ref{fig:bw_cap}a, for a 10K-user workload with a representative scaled-up HSTU-10B model,
retaining only the hottest 25\% of users
requires 3.62~TB of KV cache capacity. This far exceeds what HBM alone can practically provide, making CPU memory offloading necessary. Second, tighter latency SLOs require increasingly high KV cache bandwidth. As shown in Figure~\ref{fig:bw_cap}b, meeting a 20~ms latency target for a request with an 8K-interaction history requires 403~GB/s of KV cache read bandwidth, 
far beyond the bandwidth typically available for CPU memory. 
Together, these challenges expose a fundamental mismatch between emerging GR KV cache requirements and existing memory systems: 
HBM provides sufficient bandwidth but insufficient capacity for user-level KV caching at scale, while CPU memory offers greater capacity but is constrained by substantially lower bandwidth.\

High-Bandwidth Flash (HBF) has emerged as a promising solution to these limitations \cite{sandisk2025hbf}. By replacing HBM's DRAM with low-latency NAND flash and exploiting massive internal parallelism, HBF offers over 10$\times$ higher per-stack capacity than HBM4 (e.g., 512~GB vs. 36~GB) while approaching HBM-class read bandwidth (e.g., 1.6~TB/s per stack) \cite{sandisk2025hbf,ha2026h3,son2026hbf,kyung2026hbfkv}. These properties make HBF well suited for large, frequently accessed user KV caches. However, NAND flash has limited write endurance; typical single-level cell (SLC) flash supports only around 100K program/erase (P/E) cycles \cite{micron2026nand}. Fortunately, unlike LLM decoding, where KV caches are generated and grow within each session, GR reuses user KV caches across requests from the same user, largely decoupling KV write frequency from request frequency. Writes therefore occur mainly on cache misses, substantially reducing write traffic and making GR more amenable to HBF-based KV caching. In this work, we consider an HBF-based GR serving system that reserves one HBM4 stack for frequently updated intermediate activations and places model weights and user KV caches across seven HBF stacks to leverage their read-dominant access patterns, as shown in Figure~\ref{fig:archi}b \cite{son2026hbf}.


\begin{figure}[!t]
\centering
\includegraphics[width=0.95\columnwidth]{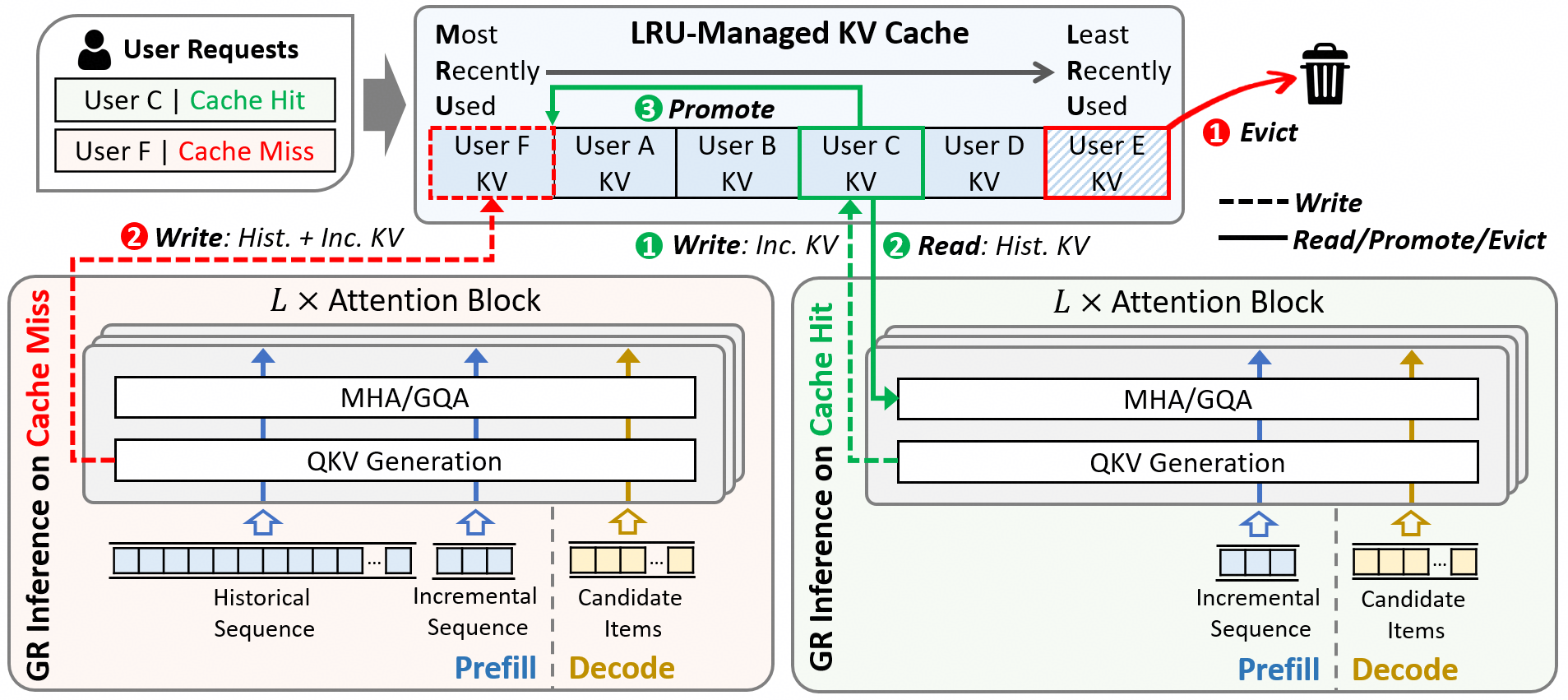}
\caption{GR Serving Workflow with Conventional LRU-Managed KV Cache. On a cache hit, historical KV is read and incremental KV is written; On a cache miss, both historical and incremental KV are written.}
\label{fig:workflow}
\end{figure}

However, despite GR's lower KV write frequency than LLM decoding, conventional GR KV cache management can still significantly undermine HBF endurance. Existing GR systems commonly adopt the Least-Recently-Used (LRU) policy for KV cache management \cite{wang2026mtserve,sun2026bat}, which tightly couples cache writes with cache misses. As illustrated in Figure~\ref{fig:workflow}, every cache miss triggers a full KV computation (historical KV + incremental KV) and inserts the full cache at the MRU position while evicting the LRU entry. Such write-on-miss behavior admits even one-off users, causing frequent cache churn and accelerated flash wear. For example, under a 10K-user HSTU-10B workload, a seven-stack HBF device can last only about one year. Hence, the existing LRU policy needs to be revised to sustainably leverage HBF's capacity benefit for GR serving.\

In this work, we evaluate a write-aware, admission-controlled LRU-$K$ policy \cite{o1993lru}, which selectively admits user KV caches based on the $K$-th request. Unlike conventional LRU, 
LRU-$K$ filters low-reuse users before admission, decoupling KV writes from cache misses and reducing unnecessary write traffic. To quantitatively assess its impacts on GR serving  performance and HBF endurance, we develop an analytical framework and evaluate LRU-$K$ across system configurations and GR models. Our contributions are three-fold:
\begin{itemize}
    \item We develop an analytical model for GR serving that characterizes KV cache access, serving performance, write traffic, and HBF lifetime. 
    \item We propose adopting LRU-$K$ in place of conventional LRU and model the reduction in write traffic enabled by write-miss decoupling.
    \item We evaluate diverse memory configurations (HBM-only, HBM+CPU, and HBF-based) and GR models (HSTU and OpenOneRec), and show that HBF-based systems achieve 3.8--4.7$\times$ higher throughput than HBM-only systems but suffer from limited lifetime under conventional LRU. With LRU-$10$, HBF lifetime extends to over six years while maintaining throughput.
\end{itemize}
 
\section{GR Inference Workload Modeling}
\label{sec:gr-model}
GR inference typically involves processing a user's interaction sequence (i.e., prefill) and scoring candidate items (i.e., decode) to generate recommendations. The interaction sequence consists of a long historical sequence (e.g., thousands of tokens) followed by a much shorter incremental sequence of newly arrived interactions (e.g., a few tokens). To avoid recomputing the long historical prefix, 
recent GR systems persist and reuse user-level KV caches across requests, leveraging conventional LRU for cache management \cite{wang2026mtserve,sun2026bat}.
Figure \ref{fig:workflow} illustrates this GR inference workflow.

\subsection{Serving Performance}
\label{subsec:performance}

Upon receiving a user request, the system first checks the KV cache for a hit or miss. On a cache hit, the historical KV cache is read from memory, while only incremental prefill and candidate decoding are computed; the new incremental KV is then appended to the cache. On a cache miss, the historical prefill is recomputed together with incremental prefill and decoding, and the resulting full KV cache is inserted into memory. Using a roofline analytical model, the hit- and miss-path latencies under conventional LRU are
\begin{equation}
\begin{aligned}
T_{\mathrm{hit}}
&=
\max\left\{
T_{\mathrm{inc+decode}}^{\mathrm{comp}},
T_{\mathrm{hist}}^{\mathrm{read}},
T_{\mathrm{inc}}^{\mathrm{write}}
\right\},
\\
T_{\mathrm{miss}}
&=
\max\left\{
T_{\mathrm{hist+inc+decode}}^{\mathrm{comp}},
T_{\mathrm{hist+inc}}^{\mathrm{write}}
\right\},
\end{aligned}
\label{eq:lru1-path-latencies}
\end{equation}
where $T^{\mathrm{comp}}$ is computed as the forward-pass FLOPs of the respective operations divided by the device peak compute throughput, and $T^{\mathrm{read}/\mathrm{write}}$ is computed as the transferred data volume divided by the read/write bandwidth\footnote{For tiered memory (e.g., HBM+CPU), our model additionally capture different tier-specific bandwidths, per-tier hit/miss behavior, and inter-tier data movement; details are omitted here for brevity.}. The average request latency and device throughput (QPS) are therefore
\begin{equation}
\begin{aligned}
\bar{T}_{\mathrm{req}}
=
p_{\mathrm{hit}}T_{\mathrm{hit}}
+
\left(1-p_{\mathrm{hit}}\right)
T_{\mathrm{miss}}, \ \
\mathrm{QPS}
=
\frac{1}{\bar{T}_{\mathrm{req}}} \ \mathrm{req/sec},
\end{aligned}
\label{eq:lru1-avg-latency-throughput}
\end{equation}
where $p_{\mathrm{hit}}$ is the hit rate of conventional LRU cache. Since a cache hit avoids expensive historical prefill, the higher hit rate enabled by the larger capacity of HBF reduces average request latency and improves serving throughput.

\subsection{KV Cache Write Traffic and HBF Lifetime}

Under conventional LRU, every cache miss writes the full historical-plus-incremental KV cache, while a cache hit writes only the incremental KV. Hence, the average write volume per request and the corresponding HBF lifetime are
\begin{equation}
\begin{aligned}
\bar{D}_{\mathrm{write,req}}
&=
p_{\mathrm{hit}}D_{\mathrm{inc}}
+
\left(1-p_{\mathrm{hit}}\right)
D_{\mathrm{hist+inc}},
\\
L_{\mathrm{HBF}}
&=
\frac{
C_{\mathrm{KV}} \cdot
E_{\mathrm{flash}}
}{
\lambda_\mathrm{wload} \cdot
\bar{D}_{\mathrm{write,req}}
}.
\end{aligned}
\label{eq:lru1-write-lifetime}
\end{equation}
Here, $D_{\mathrm{inc}}$ denotes the incremental KV cache size, $D_{\mathrm{hist+inc}}$ denotes the full KV cache size, $C_{\mathrm{KV}}$ represents the usable HBF capacity allocated to KV cache (i.e., HBF device capacity minus model weights), $E_{\mathrm{flash}}$ denotes the NAND flash write endurance measured in P/E cycles, and $\lambda_{\mathrm{wload}}$ represents the workload overall request rate derived from the evaluated user scale and request arrival distribution. Since $D_{\mathrm{hist+inc}}$ is substantially larger than $D_{\mathrm{inc}}$, writing the full KV cache on every miss can rapidly exhaust the limited write endurance of HBF. This motivates a cache policy that decouples cache writes from cache misses.

\section{Write-Aware LRU-$K$ Cache Policy}
\label{sec:lru-k-policy}
\begin{figure}[!t]
\centering
\includegraphics[width=\columnwidth]{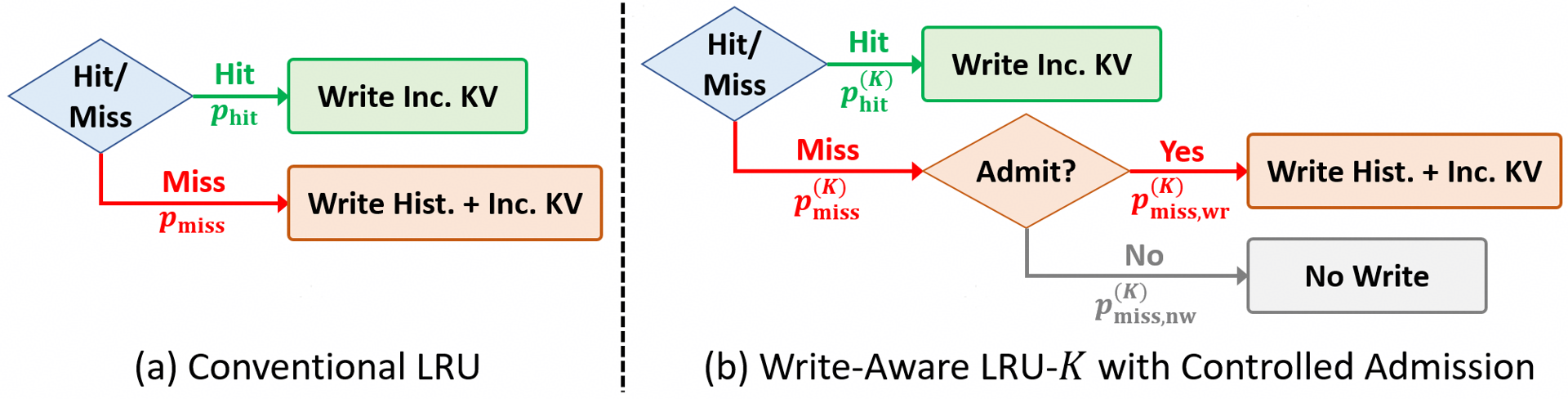}
\caption{Conventional LRU vs LRU-$K$ with Controlled Admission.}
\label{fig:admission}
\end{figure}

We consider an admission-controlled LRU-$K$ policy \cite{o1993lru}, which uses each user's $K$-th most recent request to assess their KV reuse potential for cache admission. Unlike conventional LRU, which relies on the most recent request, LRU-$K$ admits a KV cache only if \textbf{\emph{its $K$-th request is more recent than the oldest $K$-th request among cached entries}}. This criterion filters one-off or recently inactive users before persisting their KV, capturing user hotness based on both recency and frequency. As a result, KV cache writes are decoupled from cache misses, with the \textbf{\emph{miss}} path split into a \textbf{\emph{miss-write}} path and a \textbf{\emph{miss-no-write}} path, only the former of which incurs KV write traffic, as shown in Figure~\ref{fig:admission}.

\subsection{Controlled Admission and Write-Miss Decoupling}

Let
$a_K
\triangleq
\Pr\!\left(\mathrm{admit}\mid\mathrm{miss},K\right)$
denote the conditional probability that a cache miss satisfies the
LRU-$K$ admission criterion. The miss probability is therefore divided into:
\begin{equation}
\begin{aligned}
\text{\emph{miss-write:}} \quad p_{\mathrm{miss,wr}}^{(K)}
&=
p_{\mathrm{miss}}^{(K)}a_K,
\\
\text{\emph{miss-no-write:}} \quad p_{\mathrm{miss,nw}}^{(K)}
&=
p_{\mathrm{miss}}^{(K)}
\left(1-a_K\right),
\\
\end{aligned}
\label{eq:lruk-path-probabilities}
\end{equation}
where $p_{\mathrm{miss}}^{(K)} = 1-p_{\mathrm{hit}}^{(K)}$, and $p_{\mathrm{hit}}^{(K)}$ is the hit rate of LRU-$K$. Both $p_{\mathrm{hit}}^{(K)}$ and $a_K$ depend on cache capacity, request locality, and the value of $K$, and can be derived analytically or from request traces. In our evaluation, we adopt a Poisson request model, $X\sim\mathrm{Poisson}(\lambda \tau_C)$, where $\lambda$ is the request arrival rate and $\tau_C$ is the characteristic time associated with cache capacity $C$, i.e., the time window within which a user's $K$-th most recent request must occur for its KV to remain cached\footnote{Under the characteristic-time approximation, the LRU-$K$ criterion is equivalent to requiring at least $K$ requests within the characteristic time $\tau_C$.}. We can obtain $p_{\mathrm{hit}}^{(K)}=\Pr( X \geq K)=1-\sum_{x=0}^{K-1}\frac{(\lambda \tau_C)^x e^{-\lambda \tau_C}}{x!}$ and $a_K=\Pr(X=K-1\mid X<K)
=\frac{(\lambda \tau_C)^{K-1}e^{-\lambda \tau_C}/(K-1)!}
{\sum_{x=0}^{K-1}(\lambda \tau_C)^xe^{-\lambda \tau_C}/x!}$.
For LRU-1 which admits based on the most recent request, $a_1=1$, the policy degenerates to conventional LRU:
$p_{\mathrm{miss,wr}}^{(1)}=p_{\mathrm{miss}}^{(1)}=1-p_{\mathrm{hit}}^{(1)}$ and
$p_{\mathrm{miss,nw}}^{(1)}=0$.

\subsection{Impacts on HBF Lifetime and Serving Performance}

Since only hit and miss-write requests generate KV writes, the write volume per request and HBF lifetime of LRU-$K$ are
\begin{equation}
\begin{aligned}
\bar{D}_{\mathrm{write,req}}^{(K)}
&=
p_{\mathrm{hit}}^{(K)}D_{\mathrm{inc}}
+
p_{\mathrm{miss,wr}}^{(K)}
D_{\mathrm{hist+inc}},
\\
L_{\mathrm{HBF}}^{(K)}
&=
\frac{
C_{\mathrm{KV}} \cdot
E_{\mathrm{flash}}
}{
\lambda_\mathrm{wload} \cdot
\bar{D}_{\mathrm{write,req}}^{(K)}
}.
\end{aligned}
\label{eq:lruk-write-lifetime}
\end{equation}
Compared with conventional LRU (i.e., LRU-1) in \eqref{eq:lru1-write-lifetime}, LRU-$K$ reduces the full KV write rate from $1-p_{\mathrm{hit}}^{(1)}$ to $p_{\mathrm{miss,wr}}^{(K)}$.
While miss-no-write requests still incur historical prefill, they avoid writing large $D_{\mathrm{hist+inc}}$, substantially reducing write volume per request and extending HBF lifetime. 

For latency under LRU-$K$, the hit path is identical to LRU in \eqref{eq:lru1-path-latencies}, i.e., $T_{\mathrm{hit}}^{(K)}=T_{\mathrm{hit}}$, and the miss-write path matches the LRU miss path, i.e., $T_{\mathrm{miss,wr}}^{(K)}=T_{\mathrm{miss}}$. The miss-no-write path incurs only computation, i.e., $T_{\mathrm{miss,nw}}^{(K)}=T_{\mathrm{hist+inc+decode}}^{\mathrm{comp}}$. The average request latency and throughput under LRU-$K$ are
\begin{equation}
\begin{aligned}
\bar{T}_{\mathrm{req}}^{(K)}
=
p_{\mathrm{hit}}^{(K)}T_{\mathrm{hit}}^{(K)}
+
&p_{\mathrm{miss,wr}}^{(K)}
T_{\mathrm{miss,wr}}^{(K)}
+
p_{\mathrm{miss,nw}}^{(K)}
T_{\mathrm{miss,nw}}^{(K)},
\\
\mathrm{QPS}^{(K)}
&=
\frac{1}{\bar{T}_{\mathrm{req}}^{(K)}} \ \mathrm{req/sec}.
\end{aligned}
\label{eq:lruk-latency-throughput}
\end{equation}
Increasing $K$ slightly improves QPS as LRU-$K$ preferentially retains more active users who have longer histories, saving more computation on cache hits (details in Section~\ref{subsec:eval_results}).\

\section{Evaluation}
\subsection{Experimental Setup}
\paragraph{Hardware Configuration} We evaluate three memory configurations for GR serving. 1) \textbf{HBM-only}: a device equipped with eight HBM4 stacks, where each stack provides 36 GB capacity and 1.6 TB/s read/write bandwidth \cite{son2026hbf}. 2) \textbf{HBM+CPU}: augments the HBM-only design with 2~TB of CPU memory for expanded KV cache capacity, with KV cache transfers over PCIe 6.0 at 128~GB/s. We assume a write-through updates to CPU memory \cite{wang2026mtserve,sun2026bat}.
3)  \textbf{HBF-based}: a device with seven HBF stacks and one HBM4 stack. The HBM4 stack is reserved for frequently updated intermediate data.
Each HBF stack provides 512 GB capacity, 1.6 TB/s read bandwidth, and 48 GB/s write bandwidth \cite{kyung2026hbfkv}. To achieve the target read bandwidth, we assume 25 planes per HBF die (16 dies per HBF stack) and a 1-$\mu$s latency for 4-KB page reads. Following prior work \cite{son2026hbf,ha2026h3}, each HBF stack includes a 3.13-MB SRAM staging buffer on the logic die for prefetching and double buffering. We further assume HBF employs SLC flash with a write endurance of 100K P/E cycles. A peak compute throughput of 2~PFLOPs is applied for all configurations.


\paragraph{Workloads} 
We evaluate \textbf{HSTU} 1B, 10B, and a 100B MoE variant \cite{zhai2024actions}, as well as \textbf{OpenOneRec} 1.7B, 8B, and a 100B MoE variant \cite{zhou2025openonerec}. HSTU performs batched candidate scoring in a single decoding step, whereas OpenOneRec autoregressively generates item semantic tokens over few decoding steps, with each step requiring KV access. We simulate user traffic spanning diverse activity levels \cite{sun2026bat}, with hot users having request rates above 100 req/hr.
Following HSTU \cite{zhai2024actions}, we use a maximum historical sequence length of 8K, with more active users having longer sequence length. 
The candidate size is set to 500.
\subsection{Evaluation Results}
\label{subsec:eval_results}

\begin{figure}[!t]
\centering
\includegraphics[width=0.9\columnwidth]{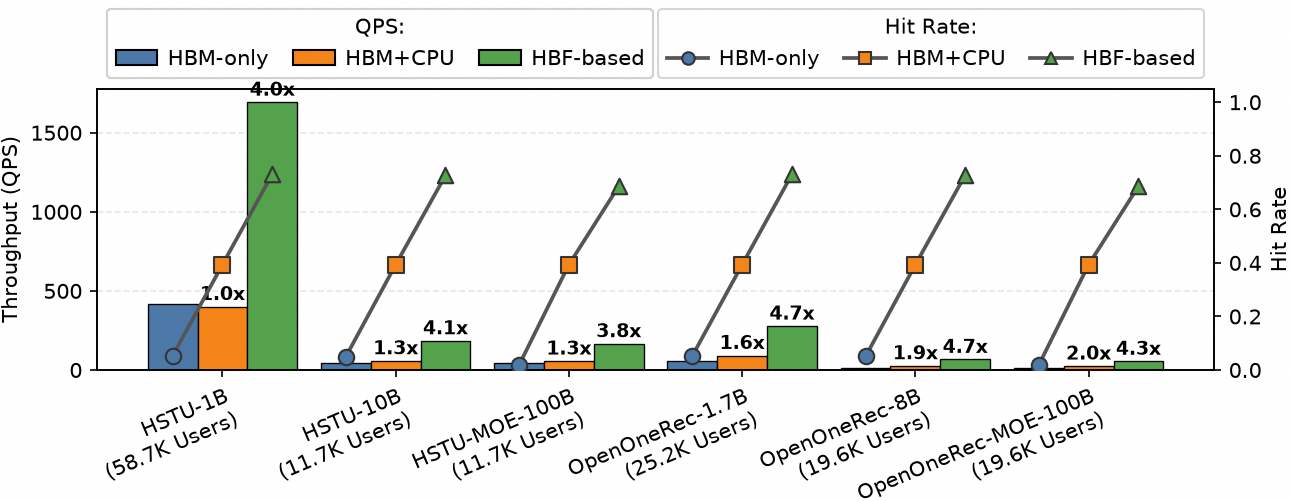}
\vspace{-2mm}
\caption{Throughput Comparison under LRU-1.}
\vspace{-2mm}
\label{fig:qps_plot}
\end{figure}

\paragraph{Throughput}
We first compare the serving throughput of HBM-only, HBM+CPU, and HBF-based configurations under conventional LRU (i.e., LRU-1). Since KV cache capacity demands vary across models, we adopt a model-specific user scale based on the capacity required to retain hot users 
in HBM+CPU. 
As shown in Figure~\ref{fig:qps_plot}, HBF-based consistently achieves higher QPS than HBM+CPU and HBM-only, mainly due to its larger KV cache capacity, which improves hit rates and reduces expensive historical prefill recomputation (Eq.~\ref{eq:lru1-path-latencies} and Eq.~\ref{eq:lru1-avg-latency-throughput}). Compared with HBM-only, HBM+CPU improves QPS by 1.3--2.0$\times$ for five of six models by expanding KV cache capacity through CPU offloading. HSTU-1B is the exception: although offloading increases its cache hit rate, the model is sufficiently small that fetching the full KV cache from bandwidth-limited CPU memory can be slower than recomputing the user history, slightly reducing throughput. With both large capacity and high read bandwidth, HBF-based configuration achieves 3.8--4.7$\times$ higher QPS than HBM-only across all models. Although an HBF stack may consume more power ($<$80W) than HBM (40W),
its higher per-device throughput can reduce the total required device count, offsetting the additional power overhead at the system level.
\begin{figure}[!t]
\centering
\includegraphics[width=0.9\columnwidth]{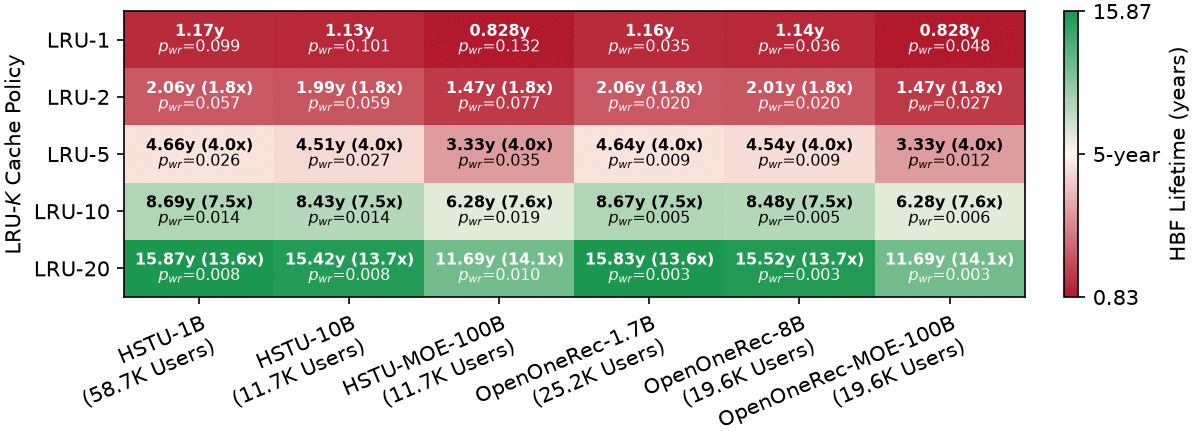}
\vspace{-2mm}
\caption{HBF Lifetime and Write Rate under Different LRU-$K$ Values.}
\vspace{-2mm}
\label{fig:lifetime_plot}
\end{figure}

\begin{figure}[!t]
\centering
\includegraphics[width=0.9\columnwidth]{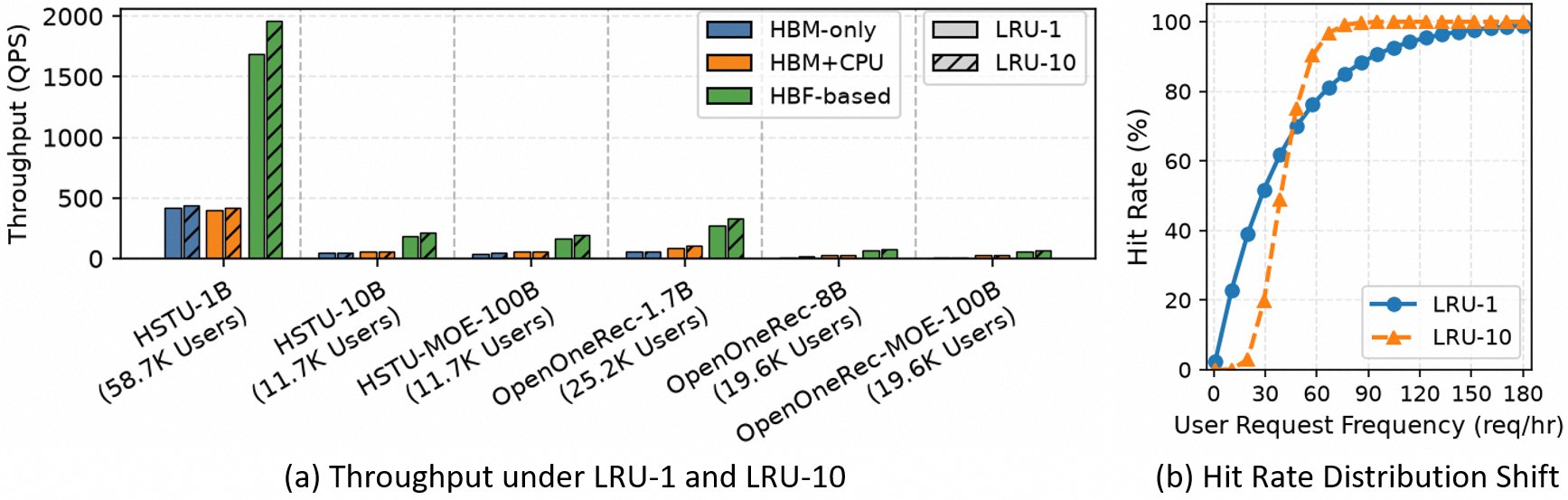}
\vspace{-2mm}
\caption{Throughput and Hit Rate Distribution Shift for LRU-1 and LRU-10.}
\vspace{-2mm}
\label{fig:qps_2_plot}
\end{figure}

\paragraph{HBF Lifetime}
For the HBF-only configuration, we evaluate HBF lifetime under LRU-$K$ with $K$ increasing from 1 to 20, where LRU-1 is equivalent to conventional LRU. As shown in Figure~\ref{fig:lifetime_plot}, LRU-1 yields an HBF lifetime of about one year. By increasing $K$ to 10, all workloads achieve lifetimes of over six years, meeting typical flash device warranty periods \cite{samsung_pm893_warranty}. This gain mainly comes from the reduction in the full KV write probability $p_{\mathrm{wr}}$. Under LRU-1, every miss triggers insertion, so $p_{\mathrm{wr}}^{(1)}=1-p_{\mathrm{hit}}^{(1)}$. In contrast, LRU-$K$ decouples cache insertions from misses, reducing the full write probability to $p_{\mathrm{wr}}^{(K)}=p_{\mathrm{miss,wr}}^{(K)}=(1-p_{\mathrm{hit}}^{(K)})a_K$. For example, for HSTU-10B, increasing $K$ from 1 to 10 reduces $p_{\mathrm{wr}}$ from 0.101 to 0.014, significantly lowering KV write traffic and extending HBF lifetime. Meanwhile, we observe that QPS slightly increases as $K$ increases from 1 to 10 (Figure~\ref{fig:qps_2_plot}a). This is because LRU-$K$ filters out infrequent users and shifts cache residency toward hotter users (Figure~\ref{fig:qps_2_plot}b). 
By retaining more active users with longer histories, LRU-$K$ avoids costly prefill of long sequences and improves serving throughput.
Thus, a moderate $K$ can substantially improve HBF endurance while maintaining or slightly improving throughput. Although LRU-$K$ involves tracking access history for admission decisions, its endurance gains outweigh the added management overhead.



\section{Conclusion}
In this work, we investigate HBF for GR serving, focusing on its write endurance limitation.
HBF offers much higher capacity than HBM and improves throughput through more KV cache retention, but conventional LRU can quickly exhaust its write endurance. We therefore evaluate an admission-controlled LRU-$K$ policy that selectively persists KV caches and decouples writes from cache misses. Across diverse GR models, HBF-based systems achieve 3.8--4.7$\times$ higher throughput than HBM-only, while LRU-$K$ extends HBF lifetime from about one year with LRU-1 to over six years with LRU-10. These results highlight the importance of write-aware KV cache policy for sustainable HBF-based GR serving.\

\bibliographystyle{IEEEtran}
\bibliography{reference}

\vfill

\end{document}